UDC 533.9

# Theoretical Analysis of Chirped Pulse Effects on Plasma Formation in Water Liquid Jet

Shireen Hilal[1], Azat O. Ismagilov[1], Anton N. Tsypkin[1], Maksim V. Melnik[1]

[1]Saint Petersburg National Research University of Information Technologies,
Mechanics and Optics, St. Petersburg, Russia

shireenhilal@itmo.ru, ismagilov.azat@itmo.ru, tsypkinan@itmo.ru, mmelnik@itmo.ru

PACS 52.38.-r

**Abstract**. We present a theoretical study of how linear chirp controls plasma density in a water jet using a two-stage framework. Stage I solves carrier-population and current equations at a single point, driven by a chirped super-Gaussian pulse. By fixing bandwidth and normalizing for intensity, we isolate a chirp-only response of plasma density, which exceeds unity and shows a consistent advantage for negative over positive chirp. Stage II propagates the field in water via the angular-spectrum method and applies the same equations across space. Normal dispersion reverses the trend: the chirp-only plasma density decreases as chirp grows, negative chirp remains less detrimental, and suppression is strongest for longer FTL pulses (e.g., 80 fs) due to dispersion-induced temporal spreading and spatio-temporal desynchronization. This study separates spectral-phase effects from bandwidth and intensity, yields testable predictions for water jets, and provides a foundation for future experiments and self-consistent propagation models.

**Key words:** Plasma density, chirped pulse.

## 1. Introduction

Plasma–liquid systems have emerged as a powerful platform for nanotechnology, offering unique physicochemical environments for the synthesis, functionalization, and assembly of nanomaterials[1–3]. When high-energy laser pulses interact with liquids, dense plasmas rich in reactive oxygen and nitrogen species (RONS) are generated [4–6], initiating reduction, nucleation, and growth processes for metallic and semiconductor nanoparticles. Unlike conventional chemical synthesis, plasma–liquid techniques operate without toxic reducing agents, allowing environmentally friendly and highly controllable nanomaterial production [7,8]. Such plasmas also enable post-synthesis surface modification, grafting functional groups onto nanostructures to enhance dispersion, catalytic activity, and biocompatibility. Applications span from catalytic nanomaterials and quantum dots for optoelectronics, to plasma-activated liquids for targeted biomedical therapies. For instance, dielectric barrier discharge plasma has been employed to functionalize graphene and carbon nanotubes, improving their dispersion in aqueous systems and enhancing their performance in energy storage and gas hydrate formation[9]. Plasma-assisted nanotechnology also provides a versatile route for producing hybrid nanostructures [10], where nanoparticles can be synthesized and simultaneously functionalized in situ, enabling direct integration into biomedical, catalytic, and sensing applications.

Within this field, laser-induced plasmas in liquids offer distinctive advantages for nanoparticle fabrication and clustering [11,12]. The high pressure, rapid quenching rates, and strong confinement at the plasma–liquid interface promote the formation of metastable phases, fine particle size control, and narrow distributions. Additionally, by tuning laser parameters such as pulse duration, beam spot size, wavelength, and focusing conditions, researchers can tailor plasma density, temperature, and lifetime, thus influencing nucleation kinetics and particle morphology[13–16].

A particularly promising approach for controlling plasma characteristics is the use of chirped laser pulses. Chirping as a temporal variation of the instantaneous frequency within an ultrashort pulse, directly modifies the rate and dynamics of energy deposition into the liquid medium [17]. Positive and negative chirps have been shown to influence ionization rates, electron heating, and plasma density, thereby enabling fine control over plasma lifetime and expansion dynamics [17,18]. While most studies on chirped pulse–plasma interactions have been conducted in gaseous or solid targets, liquid environments present unique opportunities due to their high nonlinear response, efficient ionization, and high damage thresholds compared to solids [15]. In water [17], experiments and simulations show that tuning the input quadratic chirp strongly reshapes nonlinear energy deposition and the plasma distribution near focus: under normal group velocity dispersion, negative pre-chirp drives the pulse closer to transform-limited at the nonlinear focus, increases deposited energy, and can maximize plasma length and electron density depending on the input duration. These studies used envelope propagation with Kerr and Drude terms and matched experiments by adjusting ionization and collision parameters. In argon clusters [18], nanoplasma modeling indicates a strong chirp sensitivity, ionization ignites earlier for negatively chirped pulses than for unchirped or positively chirped pulses, yet the highest peak electron density occurs for a positively chirped pulse; calculations also report ≈20% higher electron energy and larger cluster radius for negative chirp. In gas-phase wakefield acceleration [19], experiments combined with PIC simulations found a positive-chirp preference: a smoother bubble, earlier rear-bubble wavebreaking, and higher electron energies/charge than with transform-limited or negative chirp; differences appear as changes in bubble density distribution, not necessarily higher absolute plasma density for negative chirp.

Because prior liquid-phase studies typically allow peak intensity to co-vary with chirp via duration changes without normalizing it away, causally attributing changes in plasma density to chirp alone remains difficult; in our study we focus explicitly on plasma density and close this gap by (i) enforcing constant spectral width across chirp and (ii) normalizing out intensity, within a two-stage water-jet framework, single-point dynamics and water-only propagation, to quantify the intrinsic chirp effect on plasma density. All metrics, normalizations, and comparisons are defined in terms of electron-density estimates obtained from the Drude relation applied to the simulated current, evaluated across the water jet located at the focal region of focusing lens. The use of liquid jet provides a continuously renewed surface for interaction, eliminates thermal effects, and by tuning chirp parameters, it becomes possible to modulate electron density, plasma frequency, and other plasma properties. Such control can, in turn, influence nanoparticle size distribution, promote selective clustering, and enhance reproducibility in plasma–liquid synthesis processes.

## 2. Theoretical Background

Theoretical modeling in this work is based on the density matrix formalism [14,15], a framework well-suited for describing the interaction of intense ultrashort, linearly polarized optical pulses with transparent isotropic dielectric media. The model comprises three coupled equations governing the electric field $E$, plasma current density $j$, and excited-state population $\rho$.

$$\begin{cases} \dfrac{\partial E}{\partial z} - a\dfrac{\partial^3 E}{\partial \tau^3} + g\dfrac{\partial E^3}{\partial \tau} + \dfrac{2\pi}{cn_0}j = 0 \\ \dfrac{\partial j}{\partial \tau} + \dfrac{j}{\tau_c} = \beta\rho E^3 \\ \dfrac{\partial \rho}{\partial \tau} + \dfrac{\rho}{\tau_p} = \alpha E^2 \end{cases} \qquad (1)$$

Here, $n_0$ and $a$ describe normal dispersion, while $g$ represents the medium's cubic nonlinearity, related to the nonlinear index $n_2$ via $g = 2n_2/c$. $\alpha$ and $\beta$ are coefficients quantify respectively the population efficiency of highly excited electronic states and free carrier generation in the medium. $\tau_c$ and $\tau_p$ are the relaxation times for collisions and excited states. The simulation is performed in a reduced time defined by $\tau = t - zn_0/c$, where $z$ is the propagation axis. In its complete form, the first equation governs the temporal evolution of the radiation field, including the effects of plasma nonlinearities, and full field–plasma feedback. The second equation describes the dynamics of the quasi-free electron current density under the influence of the electric field, while the third governs the population of highly excited electronic states.

In the present work, we omit the first equation (field-propagation equation) from the Eq. system (1) and retain only the carrier–population and current equations because the laser waveform is prescribed as a chirped super-Gaussian (see Eq. (2)). Under this assumption the electric field acts as an external driver rather than a self-consistently evolving variable, allowing to isolate the influence of chirp parameter on electron excitation and current generation without introducing the computational complexity associated with full field–plasma coupling. This driver-only reduction is standard in breakdown modeling of liquids and dielectrics, where plasma dynamics are advanced from a known pulse profile to isolate how waveform parameters (here, chirp) control excitation and current, without the computational overhead and additional confounders of full field–plasma coupling. Representative precedents include rate-equation treatments of water breakdown driven by the laser irradiance [20], first-order multiphoton/avalanche models for ocular/aqueous media [21], and multiple-rate-equation frameworks that evolve carrier populations using a given ultrafast pulse [22], while ultrafast pulse-shaping tutorials justify imposing chirp as a predefined spectral phase.

This work employs a two-stage modeling strategy. Stage I isolates the temporal physics at a single point in space using reduced rate/current equations driven by a prescribed chirped field. Stage II constructs the spatiotemporal field $E(\tau, r, z)$ by linear focusing and dispersive propagation (angular-spectrum method), then applies the same local medium equations point-wise across $(r, z)$ to obtain $\rho(\tau, r, z)$ and $j(\tau, r, z)$. This staged approach cleanly separates the role of chirp in the temporal dynamics from beam-propagation effects before moving to fully self-consistent (envelope/UPPE) models. The input field is taken as a super-Gaussian order $m = 8$ and a linear frequency chirp implemented via a quadratic temporal phase.

$$E(\tau) = E_0 \exp\left(-\dfrac{2^{2m-1}}{t_p^{2m}}\tau^{2m}\right) \sin\left(\omega_0\left(\tau + \dfrac{A}{t_p}\tau^2\right)\right). \qquad (2)$$

where $t_p$ is the pulse duration, $\omega_0$ the carrier frequency, and $A$ dimensionless chirp parameter. Differentiating the phase shows an instantaneous angular frequency $\omega_{inst}(\tau) = \omega_0(1 + 2(A/t_p)\tau)$ a linear chirp with rate $C = 2\omega_0(A/t_p)$, A negative chirp ($A<0$) places higher frequencies at the leading edge of the pulse and lower frequencies at the trailing edge, whereas a positive chirp ($A>0$) reverses this order. We use a super-Gaussian envelope to deliberately keep the spectrum balanced, with equal energy on both sides of the pulse. In this way, chirp modifies only spectral phase without altering the energy distribution between the spectral wings as much as possible. The super-Gaussian profile also reduces long-tail artifacts and spectral sidelobes, enabling fair and reproducible comparisons between positive and negative chirp. This approach is standard practice in pulse-shaping studies aimed at isolating phase effects [23,24].

Implementing chirp inherently alters the spectral width of the pulse. In our model, to ensure a constant bandwidth across different chirp values, the duration of the Fourier-transform-limited (FTL) pulse is adjusted accordingly. However, modifying the pulse duration at fixed pulse energy inevitably changes the peak intensity. To distinguish the intrinsic influence of chirp from the secondary effects caused by intensity variation, the plasma density results were normalized by the corresponding normalized intensity values. This procedure isolates the "chirp-only" contribution to the plasma response.

## 2.1 Stage I — Single-point, chirp-driven carrier dynamics

To capture the essential plasma build-up without solving the full wave equation, we advance the density $\rho$ of highly excited electrons and the quasi-free electron current density $j$ under the driving field $E(\tau)$ via

$$\frac{\partial \rho}{\partial \tau} = \alpha E^2 - \frac{\rho}{\tau_p}, \qquad \frac{\partial j}{\partial \tau} = \beta \rho E^3 - \frac{j}{\tau_c} \qquad (3)$$

Within the Drude picture, the macroscopic current density $j$ is related to the free electron density $n_e$ and the drift velocity $v_d$ via Eq. (4) [25]:

$$\mathbf{J}(\tau) = -e n_e(\tau) \mathbf{v}_d(\tau) = \sigma \mathbf{E}(\tau), \qquad \mu = \frac{e\tau_c}{m_e}, \qquad \mathbf{v}_d = \mu \mathbf{E} \qquad (4)$$

where $e$ is the elementary charge, $\sigma$ the conductivity of the medium with μ being the electron mobility and $m_e$ is the electron mass. Combining these expressions yields an instantaneous estimate for the free electron density as:

$$n_e(\tau) = \frac{|J(\tau)|}{e|\mu E(\tau)|} \qquad (5)$$

which directly links the simulated current density from the model to the plasma density through the Drude model and Boltzmann transport framework. Because the rates scale as $E^2$ and $E^3$, the quadratic phase shifts the timing between the envelope peak and carrier phase, which modifies how $\rho$ builds before momentum relaxes on $\tau_c$. The single-point model thus isolates purely temporal chirp effects on excitation and current without beam propagation.

## 2.2 Stage II — Adding space and propagation in water

To obtain the spatiotemporal driving $E(\tau, r, z)$, we apply Fourier-optical propagation. A thin lens imposes the familiar quadratic phase $\exp[i\omega r^2/(2cF)]$ of frequency-dependent curvature. We consider a liquid water jet of $150\,\mu m$ thickness, positioned at the focal plane of a $20\,cm$ focusing lens. Since plasma formation is confined to the focal region, thus, only the jet itself is included in the numerical simulation, while the air path before and after the jet is treated analytically. Specifically, the spot size, wavefront curvature, and temporal intensity profile of the pulse at the entrance surface of the water jet are obtained from the standard Gaussian beam propagation formulas[26]. These parameters define the initial input field for the propagation inside the liquid. In this way, the simulation focuses on the medium where plasma dynamics occur, while the surrounding air regions are reduced to their optical boundary conditions. This ensures both computational efficiency and physical accuracy in describing plasma evolution within the water jet. To study plasma formation across the jet thickness, propagation over a distance $\Delta z = 150\mu m$ is carried out in the spatio-temporal frequency domain by multiplying every ($k_r, \omega$) component by the angular-spectrum propagator:

$$H(k_r, \omega) = \exp\left[-ik_z(\omega, k_r)\Delta z\right], \quad k_z = \sqrt{(n(\omega)\frac{\omega}{c})^2 - k_r^2} \tag{6}$$

Where $n(\omega)$ revers to the refractive index of the medium. $k_r$ is the radial spatial frequency, and $c$ is the speed of light. Evanescent components (when $k_r > n\omega/c$) are naturally attenuated by the propagator. This formulation is standard in Computational Fourier Optics. At each propagation step $z$, the computed field $E(\tau, r; z)$ drives the same local ODEs (1) at each radius $r$, yielding $\rho(\tau, r; z)$ and $j(\tau, r; z)$. Then maps current to plasma density by the Drude relation mentioned before. Chirp is still introduced only via the quadratic temporal phase of the input field. Within propagation, that phase translates into a different temporal ordering of frequencies along $z$, thereby altering the instantaneous plasma intensity and its overlap with the relaxation dynamics governed by $\tau_p$ and $\tau_c$ across the beam.

The present model does not feed the evolving $\rho, j$ back into the field $E$ through a refractive index change at each step; it prescribes $E(\tau, r; z)$ from linear optics and then advances the medium locally. This deliberate simplification allows us to isolate the role of chirp before introducing self-consistent propagation models such as the unidirectional pulse propagation equation (UPPE). Nonlinear contributions, including Kerr self-focusing, plasma defocusing, and the magnetic term in the Drude response, are neglected because their effect on plasma density is minimal under the studied conditions [25,27]. For example, at $I \sim 10^{13} W/cm^2$ and $\lambda = 800nm$, the electron quiver velocity give $v_d/c = 2 \times 10^{-3}$, confirming that magnetic forces are less than one percent of the electric force [28]. Similarly, Kerr and plasma nonlinearities primarily shorten the effective focal position through self-focusing or refocusing, while ponderomotive forces only redistribute already-generated electrons. Since plasma formation within the $150\,\mu m$ water jet is governed almost entirely by ionization at the local intensity, neglecting these effects does not alter the calculated electron density, aside from a small focal shift.

## 3. Results

The equation system is integrated with an explicit Runge–Kutta–Fehlberg method (ode45) on a femtosecond grid, adequate for the non-stiff dynamics here. The complete set of physical constants and pump parameters employed in the simulations is listed in Table 1.

**Table 1.** The complete set of physical constants and pump parameters used in the simulation.

| Pump parameters | |
|---|---|
| Laser wavelength, $\lambda$ | $800\ nm$ |
| Pulse energy, $E_g$ | 300 μJ |
| Focal length of the focusing lens, $F$ | 20 c$m$ |
| Beam spot diameter before the focusing lens | $2mm$ |
| Beam spot diameter at the focusing position $r$ | $\sim 102 \mu m$ |
| **Physical parameters** | |
| $\tau_c$ (fs) | $\sim 1$ [29] |
| $\tau_p$ (fs) | 150 [29] |
| $\alpha$ (SI) | $7.1687e + 19$ [29,30] |
| $\beta$ (SI) | $2.8879e + 12$ [29,30] |

Introducing a linear frequency chirp to a FTL pulse modifies its spectral width, even if the temporal envelope shape is preserved. To perform chirp-dependent comparisons at constant bandwidth, we empirically adjusted the chirped pulse duration for each chirp rate until the resulting chirped pulse spectral width reproduced the one of the FTL pulse. The adjustment was carried out by comparing numerically computed spectra (via FFT) of the FTL and chirped pulses and iteratively modifying the chirped pulse duration until the full width at half maximum (FWHM) bandwidths matched within the numerical resolution. This direct spectral matching accounts for discretization, windowing, and the super-Gaussian envelope used in our model, which can cause small deviations from ideal analytic time–bandwidth scaling. Thus, we used chirp magnitudes from $|A|=5\times 10^{-4}$ to $2\times 10^{-2}$, for FTL pulse widths of 20, 40, 60, and 80 fs. After bandwidth matching, the corresponding chirped FWHM durations at the largest $|A|=0.02$ were 20.19 fs (from 20 fs), 40.95 fs (from 40 fs), 63.45 fs (from 60 fs), and 89.49 fs (from 80 fs). This procedure ensures that, in the single-point case, any change in plasma formation reflects only the imposed chirp and the deliberately adjusted pulse duration, and, in the propagation case, the same effects with material dispersion, rather than to inadvertent differences in spectral width.

Since changing pulse duration affects the peak intensity (for fixed beam area, $I_{peak} \propto E_{pulse}/t_p$), in our simulation, we adjust the pulse duration at fixed pulse energy and isolate the chirp-only contribution to plasma density as Eq.7 represents, by dividing the normalized plasma density by the normalized pump intensity at each chirp value.

$$N_{only\ chirp} = \frac{N/N_0}{I/I_0} \qquad (7)$$

### 3.1 Results of Stage I — Single-point, chirp-driven carrier dynamics

Figure 1 illustrates how the plasma electron density depends on chirp rate for different time durations of FTL pulses. In Fig. 1a, the overall trend of the normalized plasma density is decreasing as the chirp magnitude increases, independent of whether the chirp is positive or negative. This reduction originates from the temporal stretching of the pulse: introducing chirp redistributes the spectral content of the pulse over a longer duration, lowering the peak intensity available to drive ionization.

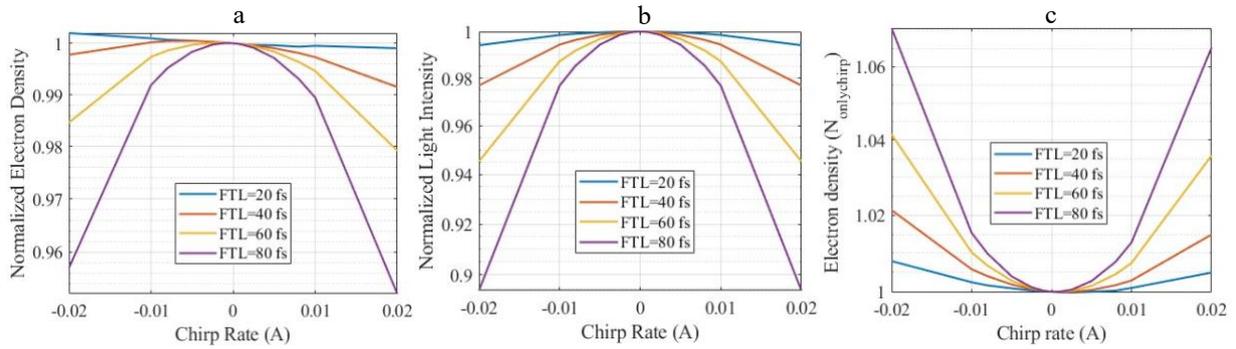

**Figure. 1.** a) Normalized plasma electron density of different FTL pulse durations at different chirp rates, b) light intensity of different FTL pulses, considering the chirped FWHM durations for different chirp rates, c) the isolated effect of introducing chirp on plasma electron density.

The corresponding normalized light intensities, adjusted to maintain a fixed spectral bandwidth, are shown in Fig. 1b. To separate the intrinsic effect of chirp from this intensity reduction, Fig. 1c presents the ratio of plasma density to normalized intensity $N_{only\ chirp}$, effectively isolating the "chirp-only" contribution. This analysis reveals that chirp has two competing influences on plasma generation. On one hand, pulse stretching lowers the peak intensity and thus reduces ionization efficiency. On the other hand, chirp alters the temporal frequency distribution within the pulse. For negative chirp, higher frequency components arrive earlier in time and coincide with higher instantaneous intensities on the leading edge of the pulse. This condition seeds ionization earlier and promotes stronger avalanche growth of free electrons. By contrast, positive chirp delays the arrival of higher frequency components until the trailing edge, where much of the available intensity has already been depleted, resulting in weaker plasma formation. The outcome is an overall asymmetry: negative chirp produces higher electron densities than positive chirp of the same magnitude. To highlight this effect, Fig. 2 provides a magnified view of the negative chirp region from Fig. 1a. At small negative chirp values, the normalized plasma density slightly exceeds unity, meaning that plasma formation is actually enhanced compared to the transform-limited case.

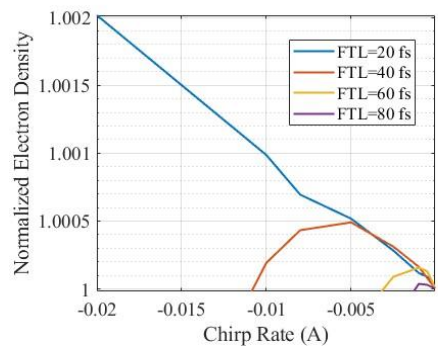

**Figure. 2.** Magnified view of the negative chirp region from Fig. 1a, showing normalized plasma electron density for different FTL pulses at various chirp rates.

This indicates that in this regime, the chirp-induced temporal ordering of frequencies enhances ionization more strongly than the intensity reduction suppresses it. Our finding matches water experiments and simulations that chirp sign matters, and suitable negative chirp reduces transmission and increases plasma density and deposited energy [17,31]. At larger chirp values, however, pulse stretching dominates and the plasma density falls below the baseline. As a result, chirp modifies plasma formation through a balance of pulse stretching (which suppresses ionization) and temporal frequency ordering (which can enhance it). Negative chirp is particularly effective because it aligns the high frequencies followed with the highest intensities at the leading part of the pulse.

### 3.2 Results of Stage II — Adding space and propagation in water

Using the same normalization procedure in Stage I (Fig. 1c), Fig. 3a. reports the chirp-only plasma response $N_{only\ chirp}$ resulting from the propagated field $E(\tau, r; z)$ in water. The curves in Fig. 3a. show chirp-only contribution for four FTL durations (20, 40, 60, 80 fs). It can be seen that all curves peak at chirp $A=0$ and decrease as $|A|$ grows, for a given $|A|$ negative chirp exceeds positive chirp where left branch lies above right, and the 80 fs FTL curve falls off most steeply, the suppression is progressively weaker for 60 fs, 40 fs, then 20 fs.

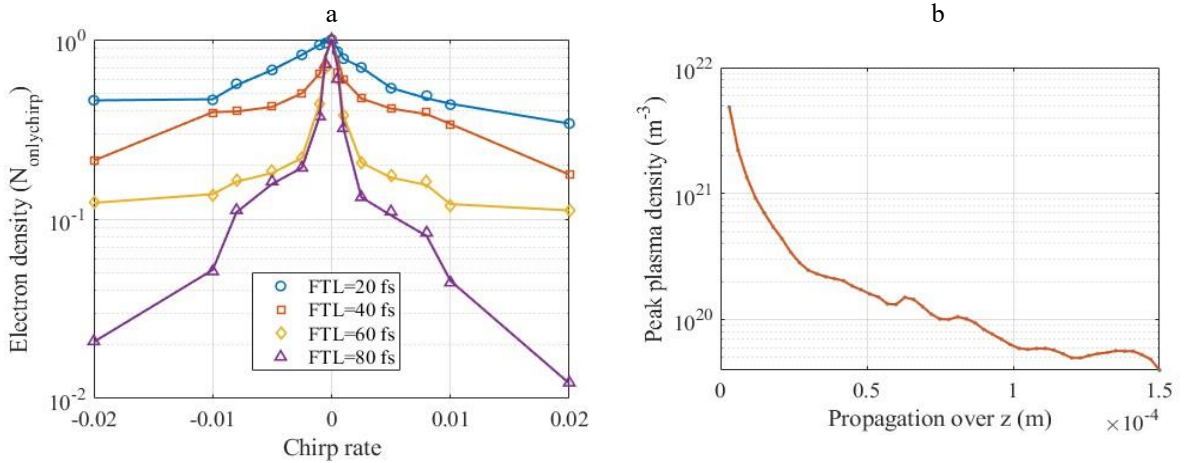

**Figure. 3.** a) Plasma electron density under the isolated effect of chirp for several FTL durations in case of propagated field $E(\tau, r; z)$ in water. b) Calculated peak plasma density per slice from the propagated field $E(\tau, r; z)$ in water as a function of propagation distance $z$, FTL duration of 20 fs with no chirp applied.

Unlike single point results in stage I, $N_{only\ chirp}$ values in Fig. 3a. fall below unity because of several factors. First, our intensity normalization equalizes input bandwidth and peak power but does not fully account for the additional pulse stretching that occurs during propagation in water, where the group-velocity dispersion at $800\ nm$ is positive and measured as $\beta_2 \approx 24.8 - 25.7\ \text{fs}^2\text{mm}^{-1}$ [32], adding a group delay dispersion of $\Phi_2 = \beta_2 z \approx 3.7 - 3.9\ \text{fs}^2$, the corresponding dimensionless chirp is $C_{disp} = 4\ln(2\Phi_2)/t_d^2$, hence the chirp added by propagation over z is $A_{disp} = C_{disp} t_p / 2\omega_0$ which gives $A_{disp} \approx 0.1104\ (20\ \text{fs})$, $0.0552\ (40\ \text{fs})$, $0.0368\ (60\ \text{fs})$, and $0.0276\ (80\ \text{fs})$. These values exceed the largest applied $|A|$ in our simulation ($\pm 0.02$). Thus, a pulse that initially carries negative chirp becomes unchirped after a compensation distance z and then acquires positive chirp for the rest of its propagation. In contrast, a pulse that begins with positive chirp gains a further increasing chirp during propagation. The associated pulse duration broadening lowers the local peak intensity across the focal region. Beyond this amplitude mechanism, dispersion also alters temporal ordering of the field while propagation, so the overlap of the medium's response ($\rho$ and the $\rho E^3$ drive that builds plasma) drops faster than an $E^2$ based intensity measure, further reducing the normalized plasma yield [33]. This effect can be seen in Fig.3b. which represents peak plasma density over the propagation distance on z with no chirp applied. The resent factors pushes $N_{only\ chirp} < 1$ even before phase-timing effects are considered [32]. Moreover, angular-spectrum propagation couples space and time since $k_z(k_r, \omega)$ varies with both frequency and transverse spatial frequency, as a result, different spectral and radial components accrue different delays and diffraction, desynchronizing on-axis and off-axis contributions [34]. Consistent with prior water studies, negative chirp still performs better than positive at equal $|A|$, but once propagation-induced dispersion and these overlap effects are included then the net chirp-only whether negative or positive response is suppressive.

## 4. Conclusion

We presented a two-stage, spectral bandwidth-controlled and intensity-normalized analysis of how linear chirp governs plasma generation in a water jet. In Stage I (single-point), intensity-normalized analysis revealed a genuine chirp-only enhancement of plasma yield, with a persistent advantage for negative over positive chirp. In Stage II (propagation in water), the trend reverses once dispersion is included: normal GVD in water at $800\ nm$ adds positive quadratic phase that stretches the pulse and degrades the timing between the field and the medium's response, so the plasma-per-intensity ratio drops below unity, yet negative chirp remains less detrimental than positive, and the suppression is strongest for longer FTL pulses. Our simulation further shows that the dispersion-induced chirp exceeds the largest applied chirp, so a pulse starting with negative chirp is quickly compensated, becomes unchirped, and then turns positive, this observation is important for experimental conditions, since commonly used media exhibit positive dispersion, and even a small thickness of $150\ \mu m$ is sufficient to cancel an initially negative chirp during the propagation. This study directly addresses common limitations in prior liquid-phase studies by holding bandwidth fixed across chirp scans and explicitly normalizing out intensity, thereby exposing the genuine effect of chirp on plasma density, independent of bandwidth and intensity. Looking ahead, the approach provides specific, testable predictions and a clean baseline for experiments, as well as a natural springboard to self-consistent propagation models to quantify chirp–dispersion balance under realistic focusing conditions. The findings provide new insight into the active control of plasma–liquid interactions and their implications for tailored nanoparticle synthesis and clustering in advanced nanotechnology applications.

## 5. Aknowledgements

This work was supported by the Ministry of Science and Higher Education of the Russian Federation (No. FSER-2025–0007)